# A SECURE METHOD FOR SIGNING IN USING QUICK RESPONSE CODES WITH MOBILE AUTHENTICATION

**Kalpesh Adhatrao[1], Aditya Gaykar[2], Rohit Jha[3], Vipul Honrao[4]**
Department of Computer Engineering, Fr. C.R.I.T., Vashi, Navi Mumbai, India
[1]kalpesh.adhatrao@gmail.com, [2]adityagaykar@gmail.com, [3]rohit305jha@gmail.com
[4]mithunhonrao2000@gmail.com

*Abstract*

*The emerging threats to user privacy over the internet are increasing at an alarming rate. Signing-in from an unreliable terminal into a web account may result in compromising private details of a user such as username and password, by means of keylogger software. Such software are capable of recording keystrokes secretly, via covert channels without the knowledge of the user. In this paper we propose a secure method for signing in using Quick Response (QR) codes with mobile authentication. Through this method, the user can securely sign-in into a web account by authenticating the user session on an unreliable terminal browser, using a mobile device.*

*Keywords-* Access tokens, Asymmetric key cryptography, CAPTCHA, Login authentication, QR code

## I. INTRODUCTION

In today's digital age, the need to secure user privacy is at its peak. There are many news reports of privacy breaches and identity thefts over the internet, compromising secret details of internet users around the world. These threats are primarily caused due to faulty login systems for web accounts on various websites, the most common being the manual password based login where the user enters a username or an email address in combination with a password. Other login techniques involve biometric authentication, online keyboards and password entry using eye tracking. These existing systems have several vulnerabilities hidden in them, which are not yet addressed to avoid possible threats. For instance, the manual method of entering the username and password combination can prove to be a disaster over an unreliable terminal, as a possibly-existent keylogger software may steal the private login credential of the user. A user is left unnoticed about this activity and is only made aware when damage is already done to his/her virtual privacy.

## II. LITERATURE SURVEY

### A. Existing login systems

### 1. Simple password-based login

It is the simplest and oldest method of entity authentication, where password is something that the claimant knows. This authentication scheme is used in the form of Fixed Password, One-Time Password or Challenge-Response. A Fixed Password is one that is used over and over for every access. A One-Time Password, as the name implies, is used only once. In this case, the user has a list of acceptable, distinct passwords, each of which can be used just once. In this way, an attacker is incapable of stealing the identity of users. The advantage of password-based login systems is that they are cheap and simple, requiring only elementary software technology.

## 2. Challenge-Response login method

In this method, authentication is done without communication of a password. Instead, a user's identity is verified based on his/her response to a challenge. The challenge is a time-varying value such as a random number or a timestamp that is sent by the verifier. The claimant applies a function to the challenge and sends the result, called the response, to the verifier. This response shows that the claimant knows the secret, which is the password to authenticate a user. The advantage of using a Challenge-Response method for login is that since there is no transfer of the password between the user and server, it becomes difficult to obtain login details by intercepting the communication information.

## 3. Virtual keyboard

Virtual Keyboards are commonly used as an on-screen input method in devices with no physical keyboard. Users input password by tapping keys of a virtual keyboard built into the device. Usually, a Virtual Keyboard has fewer buttons/keys than a normal computer keyboard. The advantage of using this method is that the susceptibility to keylogger is overcome. In order to increase safety further, the keys may be displayed in a random, shuffled order every time. This prohibits tracking of mouse movements and clicks.

## 4. Biometric authentication

Biometric authentication or Biometrics is the measurement of physiological or behavioral features that identify and authenticate an individual. These physiological and behavioral features are also called biometric identifiers. A physiological biometric authentication system identifies users using fingerprint scans, iris scans, retina scans, facial recognition, palm prints, voice recognition and DNA tests [1]. A behavioral biometric authentication system identifies users based on signatures, typing rhythm, gait and voice. The advantage of biometrics is that the information can be used to uniquely identify an individual in spite of variations in time [2].

## B. Threats and vulnerabilities in existing systems

## 1. Keylogger

A keylogger is surveillance software or spyware capable of recording and encrypting every keystroke that is made [3]. A keylogger can even record instant messages, email and any information a user types using a keyboard in a log file. This file can also be sent to a specified receiver. In this way, an attacker can obtain sensitive information from victims.

## 2. Shoulder surfing

Shoulder surfing refers to using direct observation techniques, such as looking over someone's shoulder, to get information [4]. It is commonly used to obtain passwords, PINs, security codes, and similar data [5].

## 3. Screen capturing software

Screen capturing refers to the act of copying what is currently displayed on a screen to a file or printer. If the system is in graphic mode, the screen capture will result in a graphic file containing a bitmap of the image. If the system is in text mode, the screen capture will normally load a file with ASCII codes. Screen capturing software can be used by attackers to "grab" screenshots containing login information from victims. For instance, even if a user is entering details using an online keyboard, the malicious software may take images for every click and send them to the attacker.

**4. Issues in biometric authentication**

Certain physical characteristics such as voice, gait and fingerprints may differ with time. So, it is prudent to not have authentication solely based on biometrics. Besides, the physical and behavioral characteristics of an individual are non-revocable, non-secret and thus pose a physical threat to the user [6]. Since biometrics is currently in its nascent stages, the technology is expensive and not mature yet.

**5. Accessing a password file**

An attacker can break into a system and gain access to the password file. The file may then be read by the attacker, or the contents may even be modified so that the actual users would be unable to access their accounts.

**6. Dictionary attack on passwords**

A dictionary attack uses a targeted technique of successively trying all words in an exhaustive list, called dictionary that is a pre-arranged list of values. It tries only those possibilities which are most likely to succeed, typically derived from a list of words [7]. Dictionary attacks succeed because many people have a tendency to choose passwords which are short, those with length 7 characters or less, single words found in dictionaries or simple, easily predicted variations on words, such as appending a digit.

**7. Mouse/cursor and eye tracking**

Mouse and eye tracking software have been used to study the spatial and temporal dynamics in usability testing, psychology and cognitive science, and are available for free use. Using mouse tracking software, an attacker would be able to generate images of the cursor paths and time spent at a few points to track a victim's login procedure, such as that while using on-screen and online keyboards. By measuring the point of gaze and motion of an eye relative to the head, it is possible for an attacker to determine the on-screen keys, buttons and other interactive content accessed by a victim.

**III. PROPOSED AUTHENTICATION SYSTEM**

*A. System architecture*

We propose a new login system, which strengthens the virtual privacy of a user. The objective is to provide a reliable login technique for the user, operating on an unreliable terminal, such as one in a cyber cafe. Fig. 1 describes the architecture of the entire authentication system.

In our implementation, a user's mobile phone (with an active data connection) acts as a helper device to authenticate the user on the host machine. This enables login into a web account securely without entering any private login details on the host machine. The interfacing between the host machine and the mobile phone is done using Quick Response Codes, commonly known as QR codes, and Asymmetric key technique. Two unique, random public and private pairs of access tokens, also known as asymmetric key pairs, are generated for every user login session. The private access token is stored as a session variable on web browser of the host machine. The QR code is then generated, consisting of an encoded mobile website URL (Uniform Resource Locator) along with the public access token for the user session. The data within this QR code is accessible using barcode scanner applications commonly found on smart mobile devices. The scanned URL is then used to access the mobile website through a mobile browser for authentication. This authentication initiates a trigger on the web server and causes a background update on the host machine, using the unique private key present for the user session. After the login is successful, the pair of access tokens is then discarded, and user is forwarded to access the user account. If the login is unsuccessful or the login process is terminated without completing it then the asymmetric key pair is discarded automatically after a period of 10 minutes. Since the allotted time is quite sufficient to complete the entire process.

### B. Advantages of proposed system

Our authentication system overcomes the vulnerabilities resulting due to use of keyloggers deployed by attackers covertly over unreliable terminals, since the terminal keyboard is not used at all. By shoulder surfing, an attacker can only look at the characters being entered through the keyboard and as the authentication system uses asymmetric key technique, even if some unauthorised person gets access to the QR code and scans it to gain access to the mobile website and the public key, the private key would still belong to the user session on the terminal web browser operated by the user himself. Thus users of this system are safe from shoulder surfing attacks. As one can notice, this system is also secure from screen-capturing, mouse and eye tracking software. Besides, encryption of data being communicated and also of stored passwords prevents unauthorized access and modification to information. This security is strengthened by limiting the lifetime of the access tokens being generated.

## IV. TECHNOLOGIES USED

### A. CAPTCHA

CAPTCHA stands for "Completely Automated Public Turing test to tell Computers and Humans Apart". It is a program that protects websites against bots by generating tests that humans can pass but current computer programs cannot. In our implementation, CAPTCHA is used to make sure that only humans are asking for mobile authentication [8].

### B. SHA-1

Access tokens are 10-character long pseudo-random alphanumeric keys generated through a PHP (PHP: Hypertext Preprocessor) script. This string is then encrypted using SHA-1 (Secure Hash

Algorithm - 1), which is available as a built-in function in PHP and generates a 160-bit hexadecimal string [9].

### C. QR codes

Due to its fast readability and large storage capacity compared to traditional UPC (Universal Product Codes) barcodes, the QR code has become popular for a variety of applications including storing URLs [10]. Also, apps for scanning QR codes are available on almost all smartphones and tablet PCs [11]. We have used the Google API (Application Programming Interface) to generate QR codes of access tokens.

### D. jQuery

jQuery is a fast and concise JavaScript Library that simplifies HTML document traversing, event handling, animating, and Ajax interactions for rapid web development [12]. Our implementation makes use of it to continuously ping the server to check for occurrence of a successful login by the user through a mobile phone.

### E. PHP & MySQL

PHP (recursive acronym for PHP: Hypertext Preprocessor) is a widely-used open source, general-purpose, server-side scripting language that is especially suited for web development and can be embedded into HTML [13]. It forms the front end of our implementation.

MySQL is the most popular open source Relational Database Management System (RDBMS) which is developed, distributed, and supported by Oracle Corporation. It has been used to store user profile data as well as encrypted keys and the random tokens generated during the authentication process [14].

### F. Android SDK

The Android SDK provides the API libraries and developer tools necessary to build, test, and debug apps for the Android platform [15]. We have used the Android SDK to develop an application for the Android platform, which would assist users to login in from their mobile devices.

## V. IMPLEMENTATION

The entire concept of mobile authentication has been implemented on our website http://www.compag.in. On the website login page, the user is provided with a link titled "Try mobile auth". Fig. 2 describes a Data Flow Diagram for this concept. In Fig. 3, we have shown the login procedure in the form of a Sequence Diagram. The following section illustrates the steps under implementation.

### 1. Mobile website for login authentication

Our implementation mainly involves use of user's mobile device as a helper. The mobile phone would be involved only for the job of authentication and once that is accomplished, the user is then forwarded to access the web account. But to make mobile device support login authentication, a mobile website using jQuery mobile API was developed which would enable the user to use the mobile web pages, specially designed, considering the universal mobile browser support of different smart phones, today available in the market.

### 2. CAPTCHA test

On the Desktop website, after clicking the link "Try mobile auth", the user has to pass the CAPTCHA test. For this purpose, we have used reCAPTCHA by Google, which has provision for an audio translation for people with visual disabilities. Additionally, an option to reload CAPTCHAs is available to users. After successfully completing the CAPTCHA test the user session is initiated.

### 3. Generating pair of Asymmetric access tokens and QR code

A randomization algorithm generates two unique strings having length of 10 characters. SHA algorithm is applied over these strings for encryption to get public and private access tokens respectively. Private access token is added as an argument in the page URL, whereas public access token is then set into the user session and a QR code is generated on the web page, with encoded mobile URL having the public access token.

As session starts, public access token in the user session is compared with the one containing in the QR code to check the uniqueness of the user. If some other user tries to access the same URL from different terminal, in that case user's session data will produce a mismatch for public access token as that in the QR code. This user will be considered invalid and would be shown a 401 error.

### 4. Database Manipulations

Database has a table which stores all the data related to the process of mobile authentication which includes attributes for public and private access token's hashes, User ID, date and time of authentication, and also a Boolean attribute which is used to verify success of CAPTCHA test. After successfully passing the CAPTCHA test, hashed public and private access tokens are stored in the database with their corresponding verify flag set to 1. The User ID field has a default value as 0 for that tuple which indicates mobile authentication is not yet completed.

On successful login from the mobile phone, using the mobile URL with public access token as a parameter, the User ID attribute of the tuple to which the public access token belongs, will get updated to the User ID of the user who has logged in from the mobile phone, and the tuple is marked authenticated when the value changes from 0 to 1.

### 5. Implementing instant query response via jQuery

The QR code is scanned by the user using the mobile device. The mobile website URL thus obtained is opened in a mobile browser and the user has to complete the manual login procedure of entering the email and password. Meanwhile, on the mobile device, user's terminal client browser repeatedly pings the database server after fixed interval of time using jQuery GET request API, to check whether user has been authenticated successfully on the mobile phone. Once the request is made, the corresponding tuple on the database server is traced and if it is marked authenticated, the session is again initiated for the user to access the web account. Simultaneously, the tuple from the database table is discarded and the user is then forwarded to access the homepage of the personal web account.

### 6. Mobile application for instant login authentication

The implementation using the mobile website is further extended to deploy the mobile authentication concept using a mobile application which can be installed independently on a mobile device operating on various mobile platforms. Fig. 5 and 6 show the user interface of the mobile application. The advantages of using a mobile application over using a mobile website would be that one can avoid logging in repeatedly for different mobile authentication sessions. The most important capability of a client mobile application is that it can be remember the user using local database, once he or she has logged into the application. Thus, whenever the user wishes to use mobile authentication, he or she just has to open the application and scan the QR code. The application would use values from the local mobile database to authenticate the public access token to the database server. The user is then forwarded to access the web account on the terminal browser.

Usage of this application eliminates the need for repeated logins on mobile website every time the user wishes to use mobile authentication, making it more user-friendly and time-efficient. If the user is already signed in to the website through this application and scans the QR code, his/her home page will be automatically displayed on the terminal browser. If the user logs out from the mobile application, he/she would have to enter the login details the next time after scanning the generated QR code. Fig. 5 and Fig. 6 show the mobile application interface.

## VI. FUTURE WORK

A web service of the current implementation can be developed so that various web applications can make use of it for their login mechanism. The current implementation makes use of periodic polling from client to the web server to check whether the authentication has been completed successfully from the mobile device. This method of periodic polling may overload the server when a considerably large number of user clients poll the web server at the same time which can also leads to denial of service. Thus, instead of using periodic polling, a persistent connection between server and client can be set up using HTML5's WebSocket API. Through this connection, the web server can push authentication data into client asynchronously, thereby reducing the load on the server as well as wastage of resources occurring in the case of continuous request/response paradigm of the periodic polling.

## VII. CONCLUSION

Today's standard methods for logging into websites are subject to a variety of attacks. We have presented and implemented an alternate and secure approach for entering user login details by using mobile devices such as cell phones, smartphones and tablet PCs equipped with active data connection as input devices. This method completely eliminates threats arising due to keylogger software, shoulder surfing and cursor tracking, thus securing login procedures on unknown or public computers.

**ACKNOWLEDGMENT**

We are grateful to Ms. M. Kiruthika and Ms. Smita Dange for their guidance and support.

**SCREENSHOTS**

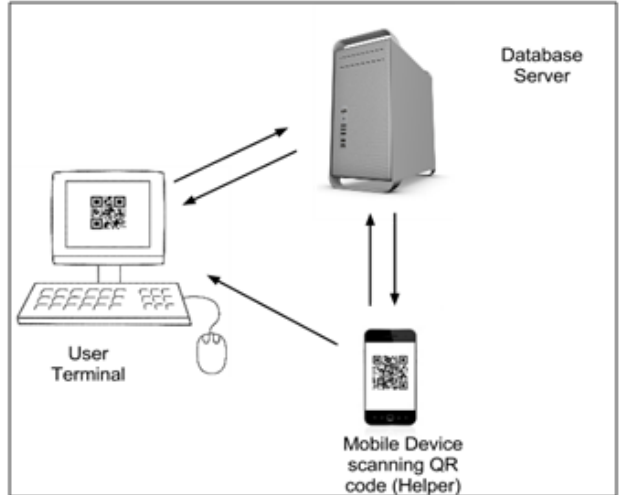

**Fig. 1. Architecture Diagram**

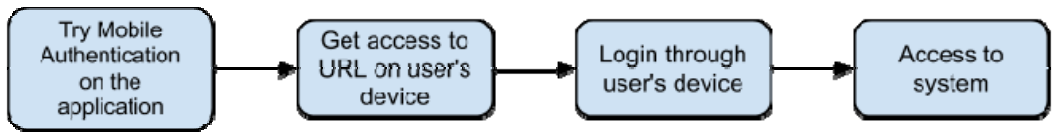

**Fig. 2. Data Flow Diagram**

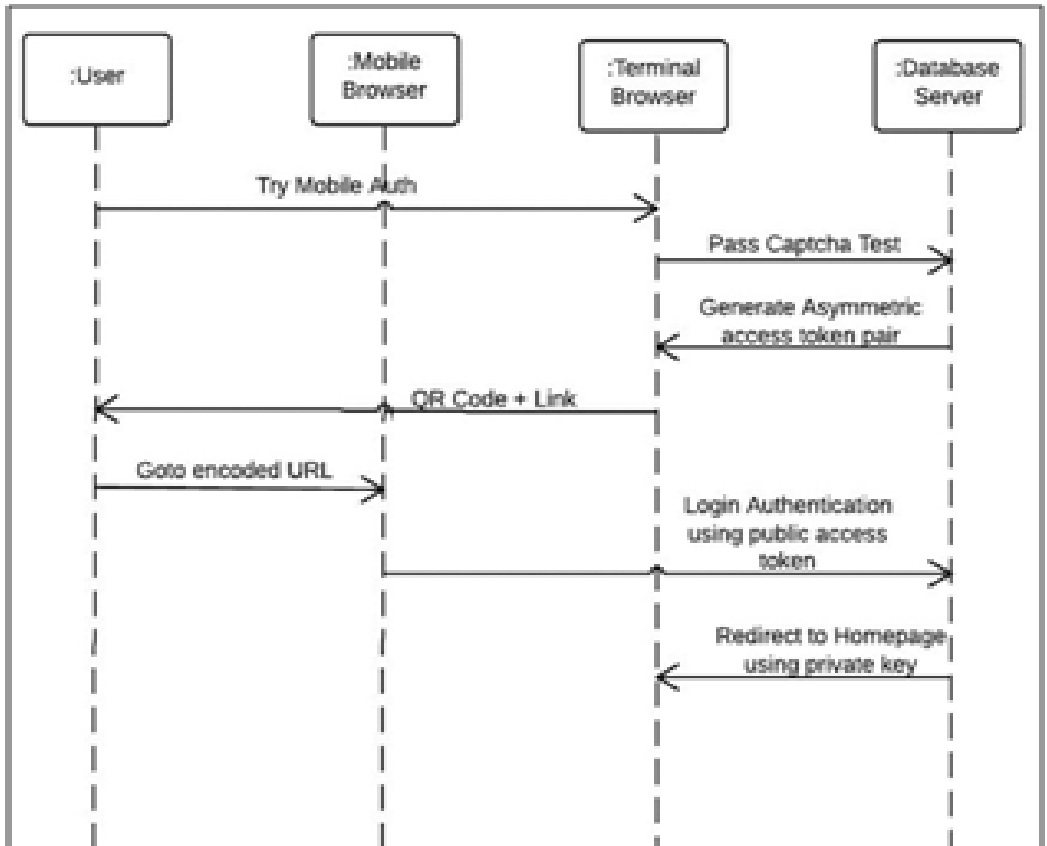

**Fig. 3. Sequence Diagram**

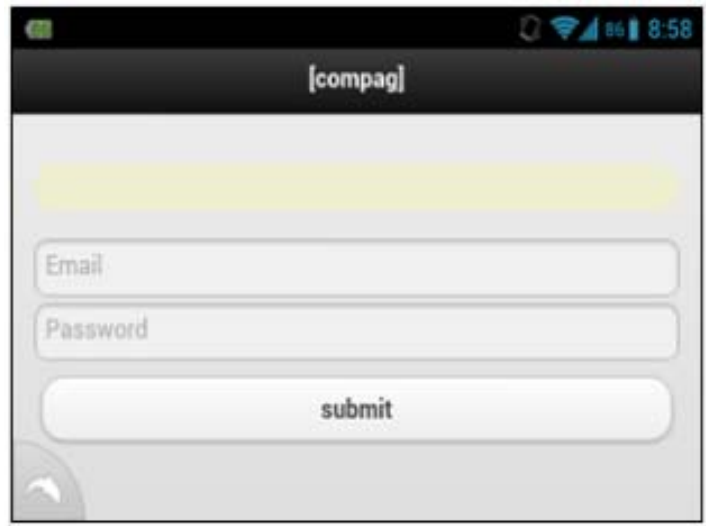

**Fig. 4. Submitting login details on the website's mobile version**

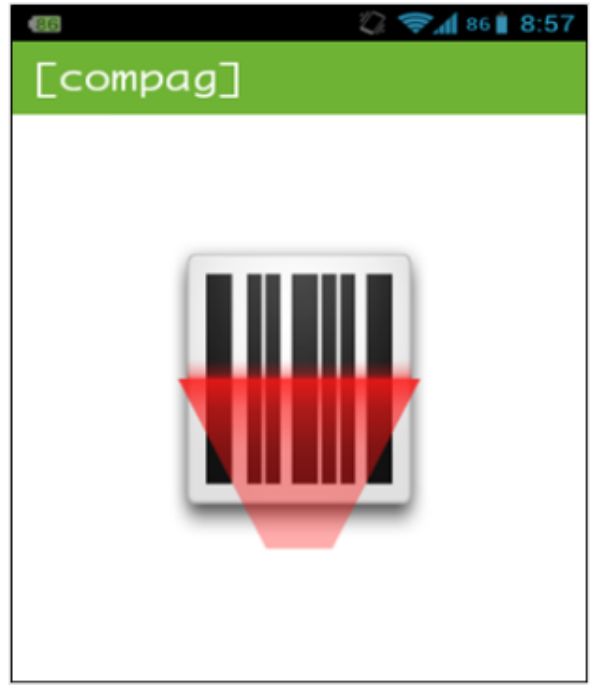

**Fig. 5. Screen for scanning QR code in the mobile application**

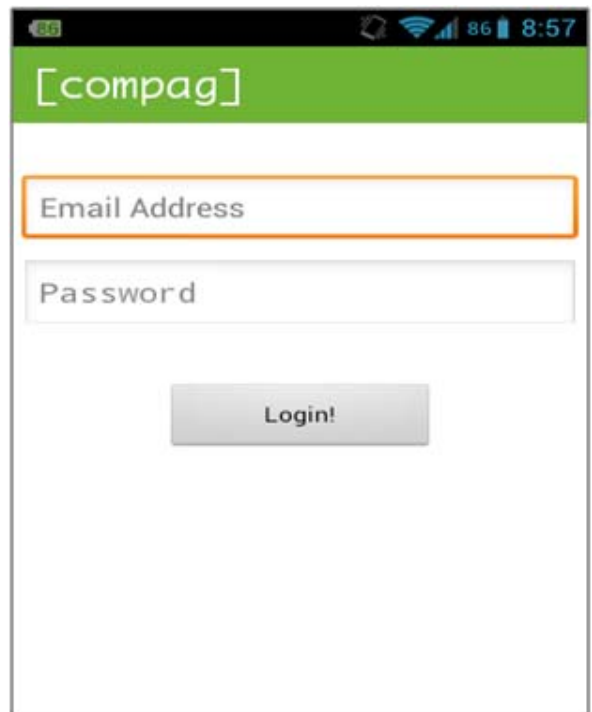

**Fig. 6. Submitting login details in the mobile application**